\documentclass{camera}
\usepackage{graphicx}
\begin{document}

\title{Time scale of the thermal \\multifragmentation in  
p(3.6 GeV) + Au collisions }

\author {\begin{quote}S.P.~Avdeyev$^1$, V.A.~Karnaukhov$^1$, H.~Oeschler$^2$, V.K.~Rodionov$^1$, 
A.V.~Simonenko$^1$, V.V.~Kirakosyan$^1$, P.A.~Rukoyatkin$^1$, 
A.~Budzanowski$^3$, W.~Karcz$^3$, I.~Skwirczy\'nska$^3$, B.~Czech$^3$, 
E.A.~Kuzmin$^4$, L.V.~Chulkov$^4$, E.~Norbeck$^5$ \and A.S.~Botvina$^6$\end{quote}}   
 
\organization{\begin{quote}$^1$ Joint Institute for Nuclear Research, 141980 Dubna, Russia \\ 
$^2$ Institute für  Kernphysik, Darmstadt University of Technology, 
64289 Darmstadt, Germany \\
$^3$ H. Niewodniczanski Institute of Nuclear Physics, 
31-342 Cracow, Poland \\
$^4$ Kurchatov Institute, 123182 Moscow, Russia \\
$^5$ University of Iowa, IowaCity,IA52242,USA \\
$^6$ Institute for Nuclear Research, 117312 Moscow, Russia \end{quote}}
\maketitle
\begin{quote}
\end{quote}

{\bf Abstract} - {\small The relative angle correlation of intermediate mass fragments has been studied for 
p+Au collisions at  3.6 GeV. Strong suppression at small angles is observed caused by 
IMF-IMF Coulomb repulsion. Experimental correlation function is compared to that obtained 
by the multi-body Coulomb trajectory calculations with the various decay time of fragmenting 
system. The combined model including the empirically modified intranuclear cascade followed 
by statistical multifragmentation was used to generate starting conditions for these calculations. 
The model dependence of the results obtained has been carefully checked. The mean decay time 
of fragmenting system is found to be $85 \pm 50$ fm/c.}      

\vspace{5mm}
{\it Introduction} - The time scale of fragment emission is a key point for understanding decay mode of 
highly excited nuclei. Is it a sequential process of independent evaporation of IMF's or 
that is a new multi-body decay mode with "simultaneous" emission of fragments governed 
by the total accessible phase space? As it was suggested in ref. \cite{sh}, "simultaneous" means 
that the primary fragments are liberated at freeze-out during the time interval which is smaller 
than the Coulomb interaction  time $\tau_c \approx 10^{-21}$s (300 - 400 fm/c). In that case 
fragment emissions are not independent as they interact via Coulomb forces while accelerating 
in the common electric field. So, measuring the IMF emission time $\tau_{em}$ ({\it i.e.} the mean 
time interval between sequential fragment emissions), or the mean life time $\tau$ of fragmenting 
system is a direct way to answer the question about the nature of the multifragmentation phenomenon. 
There is a simple relation between these two quantities via the mean IMF multiplicity \cite{li,shm}.

\qquad	Two procedures are used to determine experimentally the time scale of the process: 
analysis of the IMF-IMF correlation function in respect to the relative angle or relative 
velocity. The correlation function exhibits a minimum at $\vartheta_{rel} = 0  (\upsilon_{rel} = 0)$ 
arising from the Coulomb repulsion between the coincident fragments. The magnitude of this effect 
drastically depends on the mean emission time, since the longer the time separation of the 
fragments, the larger their space separation and the weaker the Coulomb repulsion. The time 
scale for IMF emission is estimated by comparison the measured correlation function to that 
obtained by the multibody Coulomb trajectory calculations with $\tau$ (or $\tau_{em}$) as a 
parameter. 

\qquad	The first time scale measurements for the thermal multifragmentation have been done 
in \cite{li,shm} for $^4He + Au$ collisions at 14.6 GeV by analyzing the IMF-IMF relative 
angle correlation. It was found that $\tau$ is less than 75 fm/c. Later on \cite{wa} a breakup 
time of order (20-50) fm/c was estimated via small-angle IMF-IMF relative velocity correlations 
for $^3He + Au$ interactions at 4.8 GeV and for p + Au at 8.1 GeV interaction by analyzing 
the IMF-IMF relative angle correlation is found to be $\tau \leq 70$ fm/c \cite{ro}. 
	In this paper the data on the time scale measurements for the multi-fragment emission in 
p + Au collisions at 3.6 GeV are presented. Emphasis is put on the question of the model dependence 
of the results obtained.

\vspace{5mm}
{\it Comparison between experimental data and model.} - The experiment has been performed with the $4\pi$-setup FASA \cite{av1} installed 
at the beam of the Dubna synchrophasotron-nuclotron. The device consists of two main parts: 

1)Thirty dE-E telescopes, which serve as triggers for the read-out of the system allowing 
the measurement of the fragment charge and energy distributions. The ionization chambers 
and Si(Au)-detectors are used respectively as dE and E counters.

2 )The fragment multiplicity detector (FMD) including 58 CsI(Tl) counters (with a scintillator 
thickness averaging $35 mg/cm^{-2}$), which cover $89\%$ of $4\pi$. The FMD gives the number 
of IMF's in the event and their angular distribution.

A self-supporting Au target $(1.0-1.5) mg/cm^2$ thick is located in the center of the FASA vacuum 
chamber. The beam intensity was around $7\cdot~10^8$~p/spill (spill length - 300 ms, spill period - 
10 s).

\qquad	We used a refined version of the intranuclear cascade model (INC) \cite{to,am} to get 
the distributions of the target spectators over A, Z and the excitation energy. The primary 
fragments are hot and their deexcitation is considered by SMM \cite{bon} to get the final 
distributions of cold IMF's in two break-up volume conditions: 1) freeze-out volume {\it $V_f=3V_o$}
; 2) Two characteristic break-up volumes \cite{jo}. The first volume {\it $V_t=3V_o$} corresponds 
to the stage of fragment formation, the second one {\it $V_t=5V_o$} is the freeze-out volume.

\qquad	The model calculations (INC + SMM) fail to describe the data for the IMF multiplicities 
\cite{av2,av3}. One concludes that the cascade calculation overestimates the high energy tail 
of the residue excitation energy distribution. In order to overcome this difficulty the excitation 
energies are reduced \cite{av3} event-by-event via parameter $\alpha$ :

\begin{equation}
\alpha  = \frac{<M_{exp}>}{<M_{INC+SMM}>} \label{eq:eps}
\end{equation}
					                                       
		 where $<M_{exp}>$ - measured mean multiplicities for events with at least one IMF, 
		       $<M_{INC+SMM}>$  - calculated mean multiplicities for events with at least 
one IMF. 
The mass loss during "expansion" is fine tuned via parameter $(1-\alpha)$ \cite{av3}.

\qquad	Events in model calculations (INC + $\alpha$ + SMM) with reduced excitation energies 
and reduced mass loss have been selected for IMF multiplicity $M > 2$ and at least one fragment 
has $Z > 6$. The "experimental filter" was applied to be in the line with the experimental 
definition of the correlation function. For each fragment in a given event the starting time 
to move along a Coulomb trajectory has been randomly chosen according to the decay probability 
of the system: $P(t)\sim exp(-t/\tau)$. The calculations were done for $\tau$ = 0, 100, 200 and 
300 fm/c. The left panel of fig.1. shows the comparison of the measured correlation function 
(points) with the calculated ones in case of freeze-out volume $V_f=3V_o$ for different mean 
decay times of the fragmenting system. 

\begin{figure}
\includegraphics[width=0.5\textwidth]{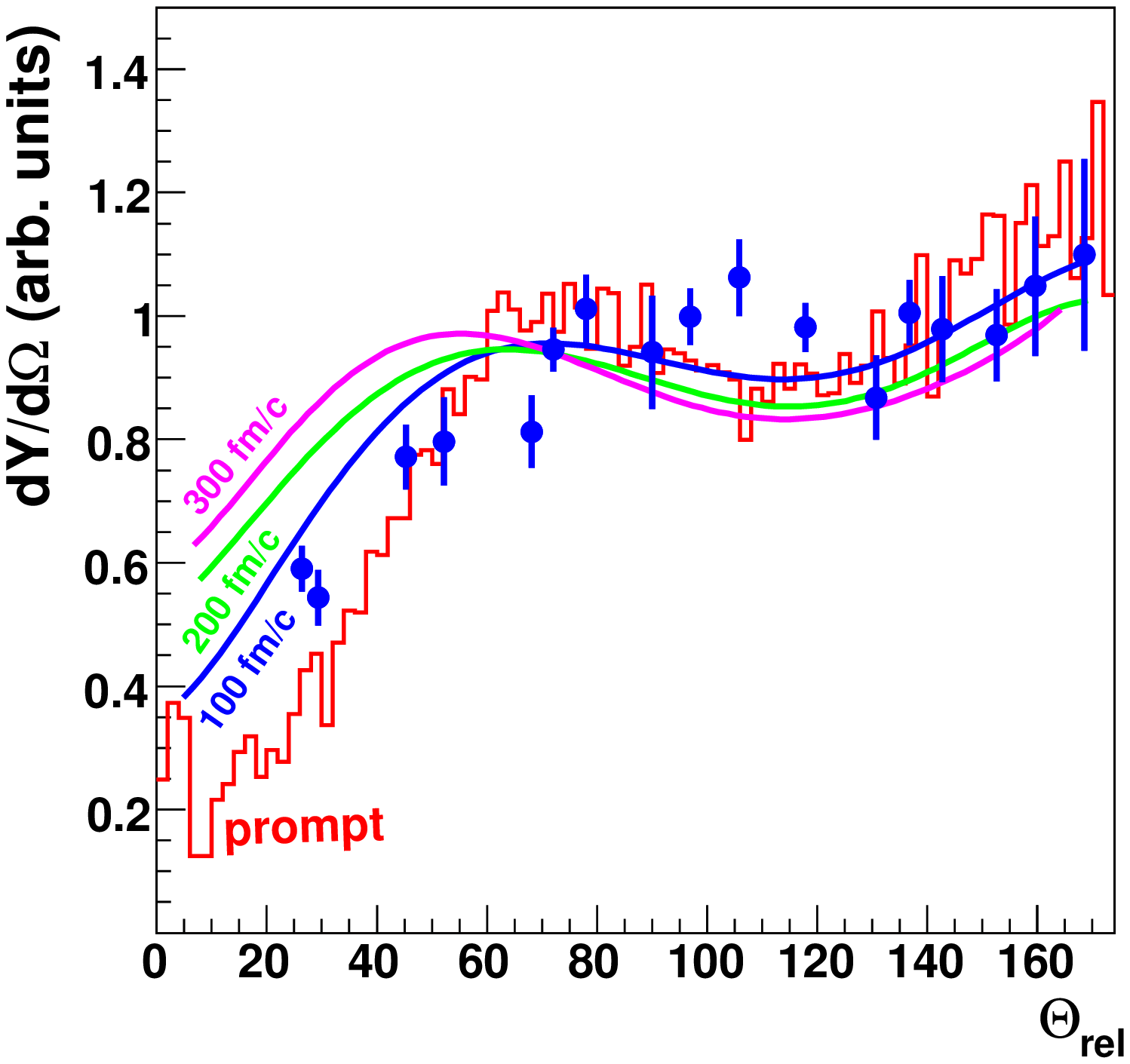}
\includegraphics[width=0.5\textwidth]{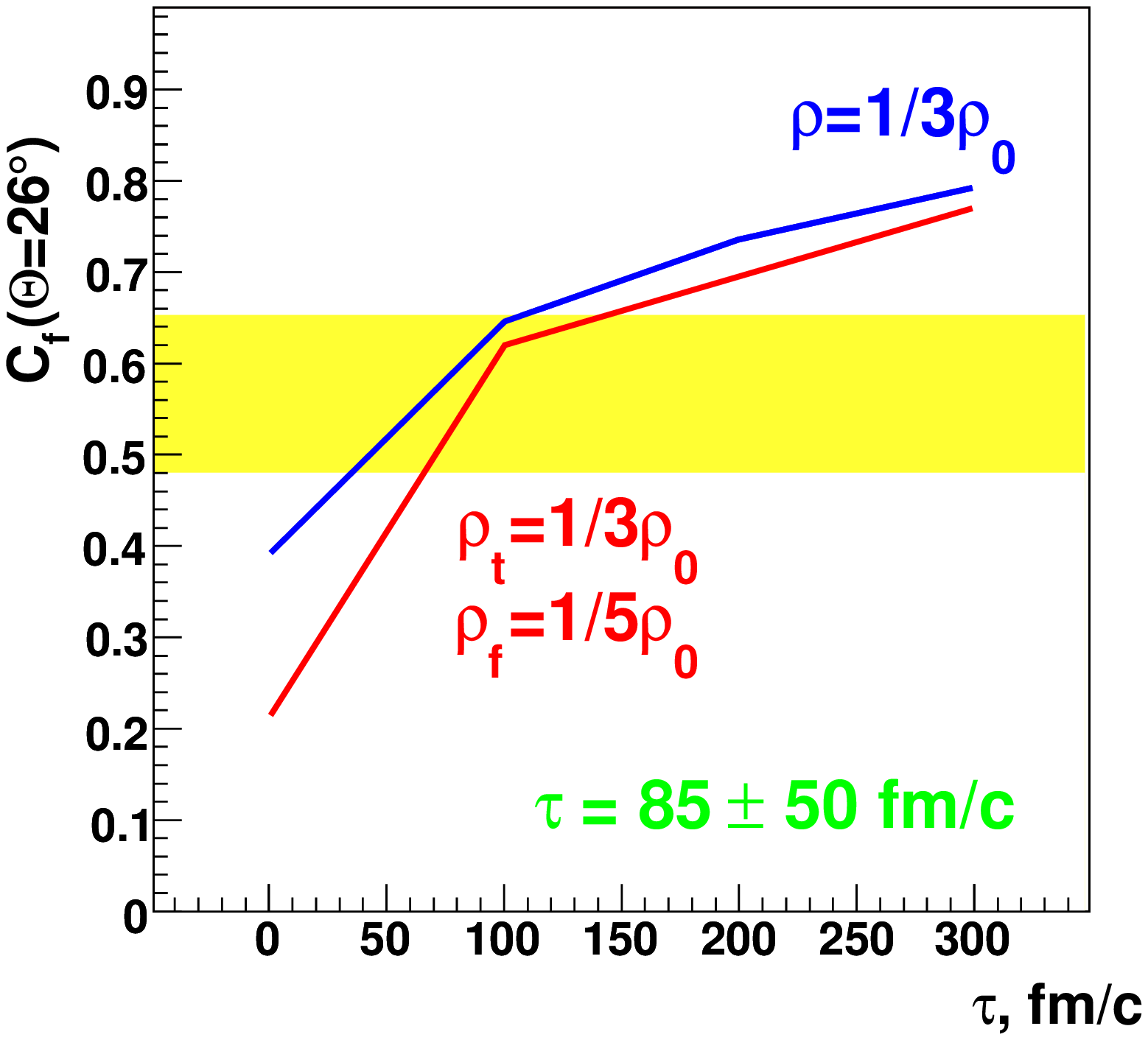}
\caption{Left panel: Relative angle correlation function for IMF produced in  p~+~Au  collisions 
at 3.6 GeV. Poins - experimental data. Histogram - INC~+~$\alpha$~+~SMM calculations with 
prompt secondary disintegration. Lines correspond to INC~+~$\alpha$~+~SMM calculations with mean 
time of secondary disintegration 100, 200 and 300 fm/c. Right panel: Correlation function at 
$\Theta_{rel} = 26^\circ $ versus the mean decay time of the system. The experimental value is 
given by the horizontal band, the lines are calculations using different decay time and break-up 
conditions.}
\label{fig.1.}
\end{figure}

\qquad Two kind of calculations have been done by the models discussed above - 
INC + $\alpha$ + SMM.  First  calculations made with freeze-out volume $V_f = 3V_o$. 
Second one used two size parameters: 

1. transition state $V_t = 3V_o$ corresponds to the stage of pre-fragment formation. 
Strong interaction between pre-fragments is still significant at this stage; 

2. freeze-out volume $V_f = 5V_o$. At this configuration, fragments are well separated 
each other, they are interacting via the Coulomb force only. 

In order to measure the IMF-IMF repulsion effect, the correlation function values at 
$\Theta_{rel} = 26^{\circ}$ is used. This quantity is shown in right panel of fig.1. 
as a function of $\tau$, the mean life time of the system. Upper line corresponds 
to calculations with freeze-out volume $V_f = 3V_o$. Lower line corresponds to 
calculations used two size parameters. The crossing of the obtained lines with the band 
corresponding to the measured correlation function and its error bar $(\pm 3\sigma)$ 
defines the mean life time of fragmenting nuclei produced in p(3.6 GeV) + Au reaction. 
The mean decay time of fragmenting system is found to be $85 \pm 50$ fm/c. 

\vspace{5mm}	
{\it Conclusion.} - The distribution of relative angles between the intermediate mass fragments 
has been measured and analyzed for thermal multifragmentation in p + Au collisions 
at 3.6 GeV. The analysis has been done on an event by event basis. The multibody 
Coulomb trajectory calculations of all charged particles have been performed starting 
with the initial break-up conditions given by the combined model with the revised 
intranuclear cascade (INC) followed by the statistical multifragmentation model. 
The distributions of the excitation energy and residual masses after INC has been 
empirically modified to reach agreement with the data for the mean IMF multiplicity. 
The correlation function was calculated for different values mean life time $\tau$ of 
the system at different break-up volume conditions, and compared with the measured 
one to find the actual time scale of the IMF emission. 

\qquad It was found good agreement of calculations and measured correlation function. 
Mean life time of the system is $85 \pm 50$ fm/c for p(3.6 GeV) + Au reaction which 
is in accordance with the scenario of a "simultaneous" multibody decay of a hot and 
expanded nuclear system.

\begin{center}
***
\end{center}     

\qquad The authors thank to A. Hrynkiewicz, A.I. Malakhov, A.G. Olchevsky for support, 
to I.N. Mishustin and W. Trautmann for illuminating discussions. The research was 
supported in part by the Russian Foundation for Basic Research, Grant No. 06-02-16068, 
the Grant of the Polish Plenipotentiary to JINR, Bundesministerium für Forschung und 
Technologie, Contract No. 06DA453.

\end{document}